\documentclass[twocolumn,tighten]{aastex61}
\usepackage[update,prepend]{epstopdf}

\newcommand\ii{{\sc ii}}
\newcommand\iii{{\sc iii}}
\newcommand\nii{[N\,\ii]}
\newcommand\cii{[C\,\ii]}
\newcommand\siii{[S\,\iii]}
\shorttitle{Origins of [CII] Emission}
\shortauthors{Croxall et al.}
\begin{document}  

\title{The Origins of [C\,II] Emission in Local Star-forming Galaxies}

\correspondingauthor{J.D. Smith}
\email{jd.smith@utoledo.edu}

\author{K.V. Croxall}
\affiliation{Department of Astronomy, The Ohio State University, 4051 McPherson Laboratory, 140 W 18th Ave., Columbus, OH, 43210}
\affiliation{Max-Planck-Institut f¬ur Astronomie, K\"{o}nigstuhl 17, D-69117 Heidelberg, Germany}
\affiliation{Illumination Works LLC, 5550 Blazer Parkway, Suite 150, Dublin, OH 43017}

\author{J.\,D. Smith}
\affiliation{Department of Physics \& Astronomy, University of Toledo, 2801 W Bancroft St, Toledo, OH 43606}
\affiliation{Max-Planck-Institut f¬ur Astronomie, K\"{o}nigstuhl 17, D-69117 Heidelberg, Germany}

\author{E. Pellegrini}
\affiliation{Universit\"{a}t Heidelberg, Zentrum f\"{u}r Astronomie, Institut f\"{u}r Theoretische Astrophysik, Albert-Ueberle-Stra{\ss}e 2, D-69120 Heidelberg, Germany}
\affiliation{Department of Physics \& Astronomy, University of Toledo, 2801 W Bancroft St, Toledo, OH 43606}

\author{B. Groves}
\affiliation{Research School of Astronomy \& Astrophysics, Australian National University, Cotter Road, Weston, ACT 2611, Australia}

\author{A. Bolatto}
\affiliation{Department of Astronomy, University of Maryland, College Park, MD 20742, USA}

\author{R. Herrera-Camus}
\affiliation{Max-Planck-Institut f\"{u}r extraterrestrische Physik, Giessen-bachstr., 85748 Garching, Germany}

\author{K.\,M. Sandstrom}
\affiliation{Center for Astrophysics and Space Sciences, Department of Physics, University of California, San Diego, 9500 Gilman Drive, La Jolla, CA 92093, USA}

\author{B. Draine}
\affiliation{Department of Astrophysical Sciences, Princeton University, Princeton, NJ 08544, USA}

\author{M.\,G. Wolfire}
\affiliation{Department of Astronomy, University of Maryland, College Park, MD 20742, USA}

\author{L. Armus}
\affiliation{Spitzer Science Center, California Institute of Technology, MC 314-6, Pasadena, CA 91125, USA}

\author{M. Boquien}
\affiliation{Unidad de Astronom\'{i}a, Fac. Cs. B\'{a}sicas, Universidad de Antofagasta, Avda. U. de Antofagasta 02800, Antofagasta}

\author{B. Brandl}
\affiliation{Leiden Observatory, Leiden University, P.O. Box 9513, 2300 RA Leiden, The Netherlands}
\affiliation{Delft University of Technology, Faculty of Aerospace Engineering, Kluyverweg 1, 2629 HS Delft, The Netherlands}

\author{D. Dale}
\affiliation{Department of Physics and Astronomy, University of Wyoming, Laramie, WY 82071, USA}

\author{M. Galametz}
\affiliation{Laboratoire AIM-Paris-Saclay, CEA/DSM/Irfu - CNRS - Université Paris Diderot, CEA-Saclay, 91191, Gif-sur-Yvette}
\affiliation{European Southern Observatory, Karl Schwarzschild Strasse 2, D-85748 Garching, Germany}

\author{L. Hunt}
\affiliation{INAF-Osservatorio Astrofisico di Arcetri, Largo E. Fermi 5, 50125, Firenze, Italy}

\author{R. Kennicutt, Jr.}
\affiliation{Institute of Astronomy, University of Cambridge, Madingley Road, Cambridge, CB3 0HA, UK}

\author{K. Kreckel}
\affiliation{Max-Planck-Institut f¬ur Astronomie, K\"{o}nigstuhl 17, D-69117 Heidelberg, Germany}

\author{D. Rigopoulou}
\affiliation{Department of Physics, University of Oxford, Keble Road, Oxford OX1 3RH, UK}

\author{P. van der Werf}
\affiliation{Leiden Observatory, Leiden University, P.O. Box 9513, 2300 RA Leiden, The Netherlands}

\author{C. Wilson}
\affiliation{Department of Physics \& Astronomy, McMaster University, 1280 Main St. W., Hamilton, ON, L8S 4M1, Canada}

\begin{abstract}
  The \cii\ $158\,\mu$m fine-structure line is the brightest emission line observed in local star-forming galaxies.  As a major coolant of the gas-phase interstellar medium, \cii\ balances the heating, including that due to far-ultraviolet photons, which heat the gas via the photoelectric effect.  However, the origin of \cii\ emission remains unclear, because C$^+$ can be found in multiple phases of the interstellar medium.  Here we measure the fractions of \cii\ emission originating in the ionized and neutral gas phases of a sample of nearby galaxies.  We use the \nii\,205\,$\mu$m fine-structure line to trace the ionized medium, thereby eliminating the strong density dependence that exists in the ratio of \cii/\nii\,122\,$\mu$m.  Using the FIR \cii\ and \nii\ emission detected by the KINGFISH and Beyond the Peak {\it Herschel} programs, we show that 60--80\% of [C~\ii] emission originates from neutral gas.  We find that the fraction of \cii\ originating in the neutral medium has a weak dependence on dust temperature and the surface density of star formation, and a stronger dependence on the gas-phase metallicity.  In metal-rich environments, the relatively cooler ionized gas makes substantially \emph{larger} contributions to total \cii\ emission than at low abundance, contrary to prior expectations.  Approximate calibrations of this metallicity trend are provided.
\end{abstract} 
 
 \keywords{galaxies: ISM --- ISM: lines and bands}
 

\section{Introduction}
The far-IR [C\,\ii]\,158\,$\mu$m fine-structure line is the strongest emission line detected in star-forming galaxies \citep{stacey1985,stacey1991,luhman1998}.   The usefulness of this line comes from the fact that (1) carbon is relatively abundant, (2) it has a low ionization potential (11.26 eV), and (3) this particular transition is relatively easy to excite ($\Delta$E/k $\sim$92 K).  As a result, all but the coldest and hottest gas is cooled by this collisionally excited line.  Due to its prominence in the far-IR spectrum, [C\,\ii] emission has been used to explore physical processes occurring in the interstellar medium (ISM) of nearby galaxies.  For example, [C\,\ii] emission has been used to trace the efficiency of heating photon-dominated regions (PDRs) via the photoelectric effect \citep[e.g.,][]{croxall2012}, used as a calorimetric tracer of the current star formation rate \citep{stacey1991, delooze2014, magdis2014,  HerreraCamus2015}, and to diagnose the thermal pressure and ionization fraction of the neutral ISM \citep[e.g.,][]{madden1997, wolfire1990,wolfire1995,beirao2010}.  Furthermore, at high redshift this line becomes one of the principal ways to trace the ISM and early star formation \citep{iono2006,wagg2010,stacey2010,carilli2013,debreuck2014,gullberg2015}.

Unfortunately, the usefulness of [C\,\ii] emission as a diagnostic is also hampered by its ubiquity.  The total integrated [C\,\ii] emission along a line of sight includes contributions from warm ionized gas, diffuse atomic and molecular gas, and PDRs near hot stars and on the surfaces of molecular clouds \citep{bennett1994,Heiles1994}.  Thus, when studying a particular phase of the ISM, contamination from emission originating in the other phases must be removed for an accurate assessment of conditions in the ISM.  For example, the PDR models of \citet{kaufman2006} only account for emission from neutral gas and thus additional [C\,\ii] emission from the ionized phase must be removed to derive a density and the incident UV flux, G$_0$, from [C\,\ii] observations.  Furthermore, high densities of star-formation can lead to a [C\,\ii] ``deficit'' that complicates understanding the origins of [C\,\ii] emission \citep[e.g.,][]{luhman1998,diazsantos2014,smith2016}.

\begin{deluxetable*}{l l l l l l l l l}[t]
\tablecaption{Region Data --- Full Table available online\label{tab:data}}
\tabletypesize{\small}
\tablehead{
\colhead{Galaxy} &
\colhead{R.A.} &
\colhead{Dec.} &
\colhead{ $\nu$f$_{\nu}$(70)/$\nu$f$_{\nu}$(160)} &
\colhead{[C\,II] 158\,$\mu$m} &
\colhead{[N\,II] 122\,$\mu$m} &
\colhead{[N\,II] 205\,$\mu$m} &
\colhead{n$_{e}$} &
\colhead{f$_{[C\,II],Neutral}$}  \\
\colhead{} &
\colhead{(J2000)} &
\colhead{(J2000)} &
\colhead{} &
\colhead{10$^{-8}$\,W\,m$^{-2}$\,sr$^{-1}$} &
\colhead{10$^{-9}$\,W\,m$^{-2}$\,sr$^{-1}$} &
\colhead{10$^{-9}$\,W\,m$^{-2}$\,sr$^{-1}$} &
\colhead{cm$^{-3}$} &
\colhead{}
}
\startdata
NGC\,1097	&	41.5796	&	-30.2829	&	1.47	$\pm$	0.40&	2.64	 $\pm$ 	0.13	&	3.37	 $\pm$ 	0.52	&	1.65	 $\pm$ 	0.49	&	53	&	0.784	 $\pm$ 	0.066	\\
NGC\,1097	&	41.5736	&	-30.2677	&	1.33	$\pm$	0.15&	2.75	 $\pm$ 	0.13	&	0.95	 $\pm$ 	1.08	&	1.53	 $\pm$ 	0.39	&	$\leq$10&	0.754	 $\pm$ 	0.061	\\
NGC\,1097	&	41.5770	&	-30.2794	&	1.90	$\pm$	0.40&	12.70	 $\pm$ 	0.13	&	19.96	 $\pm$ 	1.04	&	8.28	 $\pm$ 	0.61	&	71	&	0.770	 $\pm$ 	0.014	\\
NGC\,1097	&	41.5741	&	-30.2756	&	1.89	$\pm$	0.29&	14.62	 $\pm$ 	0.13	&	23.89	 $\pm$ 	1.35	&	12.30	 $\pm$ 	0.76	&	49	&	0.710	 $\pm$ 	0.012	\\
NGC\,1097	&	41.5768	&	-30.2710	&	2.24	$\pm$	0.51&	21.26	 $\pm$ 	0.13	&	39.29	 $\pm$ 	1.42	&	15.60	 $\pm$ 	0.53	&	77	&	0.740	 $\pm$ 	0.008	\\
NGC\,1097	&	41.5791	&	-30.2751	&	2.11	$\pm$	0.64&	45.25	 $\pm$ 	0.13	&	109.96	 $\pm$ 	3.09	&	32.20	 $\pm$ 	1.06	&	129	&	0.731	 $\pm$ 	0.004	\\
NGC\,1266	&	49.0028	&	-2.42728	&	3.32	$\pm$	0.97&	5.24	 $\pm$ 	0.13	&	5.41	 $\pm$ 	1.70	&	1.65	 $\pm$ 	0.49	&	120	&	0.882	 $\pm$ 	0.035	\\
NGC\,1482	&	58.6698	&	-20.5020	&	1.57	$\pm$	0.23&	3.49	 $\pm$ 	0.13	&	1.94	 $\pm$ 	1.25	&	1.53	 $\pm$ 	0.50	&	21	&	0.845	 $\pm$ 	0.051	\\
NGC\,1482	&	58.6612	&	-20.4978	&	2.30	$\pm$	0.56&	9.38	 $\pm$ 	0.13	&	9.11	 $\pm$ 	1.71	&	1.75	 $\pm$ 	0.51	&	289	&	0.919	 $\pm$ 	0.020	\\
NGC\,1482	&	58.6621	&	-20.5025	&	2.76	$\pm$	0.73&	58.70	 $\pm$ 	0.13	&	78.50	 $\pm$ 	1.67	&	26.60	 $\pm$ 	0.82	&	100	&	0.834	 $\pm$ 	0.003	\\
NGC\,2798	&	139.345	&	 41.9997	&	3.18	$\pm$	0.42&	24.29	 $\pm$ 	0.13	&	27.18	 $\pm$ 	1.67	&	8.13	 $\pm$ 	0.40	&	124	&	0.874	 $\pm$ 	0.007	\\
\enddata
\end{deluxetable*}

The difficulty of disentangling the origins of [C\,\ii] emission were recently highlighted by \citet{Velusamy2014} who used the spectrally resolved data from the Galactic Observations of Terahertz C+ (GOT C+)\footnote{ftp://hsa.esac.esa.int/URD\_rep/GOT\_Cplus/} Herschel Open Time Key Program to trace the origin of [C\,\ii]\,158\,$\mu$m emission in the Milky Way.  Combining the GOT C+ data with line maps from  [\ion{C}{1}], \ion{H}{1}, $^{12}$CO, $^{13}$CO, and C$^{18}$O, \citet{Velusamy2014} were able to disentangle the origins of the [C\,\ii] emission along several lines-of-sight, finding that for the Milky Way \cii\ emission primarily originates in molecular gas ($\sim$62\,\%) with the remainder coming from \ion{H}{1} gas and the warm ionized medium.  Unfortunately, applying this multi-tracer approch to study the origin of CII in distant galaxies is complicated as the spatial resolution of observations is limited, preventing the separation of different spatial components.

It is important to note that this technique can only separate phases which have distinct velocities.  \citet{Velusamy2014} designate the C$^+$ which shares a line-of-sight velocity with CO emission as originating within molecular clouds. However that emission, at the same velocity, could arise from either a skin of molecular (H$_2$) and atomic gas in a PDR or indeed from ionized gas in the adjacent H\ii\ region, limiting what can be inferred about the origin of \cii\ in ionized regions.

In a simplified model of the ISM, we can attribute the origin of [C\,\ii] emission to either ionized or neutral gas, as carbon has an ionization potential of 11.3\,eV, slightly lower than that of hydrogen, 13.6\,eV.  Conversely, \nii\ is only produced in the ionized medium as the ionization potential of nitrogen is 14.53\,eV.  We can thus trace the [C\,\ii] contribution from gas dominated by ionized hydrogen via its association with \nii\ emission \citep{bennett1994}.  Attempts to disentangle the origins of [C\,\ii] emission using observations of the [N\,\ii]\,122\,$\mu$m lines have been made \citep[e.g.,][]{kaufman1999, malhotra2001, vasta2010, croxall2012}.  However, the [N\,\ii]\,122\,$\mu$m line has a critical density for collisions with electrons ($\sim$300\,cm$^{-3}$) substantially higher than that of the [C\,\ii]\,158\,$\mu$m line ($\sim$45\,cm$^{-3}$). This leads to a strong dependence of the [C\,\ii]/[N\,\ii]\,122\,$\mu$m ratio associated with ionized gas on the gas density.  

Measuring the fraction of [C\,\ii] originating from ionized gas is more straightforward using the [N\,\ii]\,205\,$\mu$m line \citep{parkin2013,hughes2015}.  Even though this line is significantly fainter than [N\,\ii]\,122\,$\mu$m it has a critical density ($\sim$32\,cm$^{-3}$) that is similar to that of the [C\,\ii]\,158\,$\mu$m line in ionized gas \citep[e.g.,][]{oberst2006}.  This leads to a negligible dependence of the line ratio on density.  While this line proved difficult to detect in all but the brightest Herschel/PACS spectral-line observations, due to declining detector sensitivity past 200 $\mu$m, it is the brightest single line visible in SPIRE--FTS spectra of nearby galaxies \citep{kamenetzky2014,HerreraCamus2015}.

In this paper we combine measurements of the [C\,\ii]\,158\,$\mu$m and [N\,\ii]\,122\,$\mu$m lines from the KINGFISH {\it Herschel} Open Time Key Project \citep[Key Insights on Nearby Galaxies: a Far- Infrared Survey with Herschel, ][]{kennicutt2011} with measurements of the [N\,\ii] 205\,$\mu$m line from the associated project, Beyond the Peak (BtP; OT1\_jsmith1; P.I. J.D. Smith), which used SPIRE--FTS to map a subsample of KINGFISH galaxies, to deduce the fraction of [C\,\ii] emission that originates in ionized gas.  This Letter is structured as follows: Section 2 describes our {\it Herschel} observations of the fine-structure lines. In Section 3 we discuss the line ratios and possible complications in measuring the fraction of \cii\ emission arising from different phases.  Finally, in Section 4 we discuss the origin of the \cii\ fine-structure emission. 

\section{Data}  

Our observations include photometric and spectral-line observations from both the PACS and SPIRE instruments onboard {\it Herschel}; obtained as part of the large KINGFISH and BtP projects.  The overlap between these programs consists of 21 central regions and 2 extra-nuclear regions that were initially selected from the SINGS Survey \citep{kennicutt2003}.  Basic galaxy properties were taken from \citet{kennicutt2011}.  Oxygen abundances were taken from \cite{moustakas2010}.  For galaxies where \cite{moustakas2010} provides a metallicity gradient, we have de-projected the galaxies and assigned metallicities of individual regions using the optical radius, inclination and position angles provided in \citet{hunt2015}.  We note that this does assume that galaxies do not have azimuthal scatter and that the abundances across an individual field (16\farcs8) are uniform.  

\subsection{SPIRE spectroscopy}
BtP observations were performed with SPIRE--FTS intermediate mapping, a 4-point dither.  A brief description of the data can be found in \citet{pellegrini2013}. Pellegrini et al. (2017, in prep.) will contain a full description of the observations and reductions.  We note that the telescope response function was calculated from the extended catalog of darks obtained late in {\it Herschel's} mission. Using these darks we have subtracted a high-order noiseless polynomial fit of the dark, bolometer by bolometer.  We have used the individual bolometer fluxes from 281 sub-regions distributed across the 23 targeted regions, after applying an extended source correction \citep[see][for additional details]{HerreraCamus2016}.  Line fluxes and the associated uncertainties are listed in Table~\ref{tab:data}.

\subsection{PACS photometry \& spectroscopy}
[C\,\ii] and [N\,\ii]\,122\,$\mu$m emission lines, and 70 and 160 $\mu$m continuum maps have been observed as part the {\it Herschel} open time key program KINGFISH. The data and the associated reduction are described in \citet{kennicutt2011} and \citet{croxall2013}.  We have extracted line and continuum fluxes from regions corresponding to the positions and sizes of the SPIRE--FTS bolometers.   Bolometers which did not fully overlap the footprint of the [C\ii] maps were eliminated from our sample.  Before extraction, maps were convolved to match the resolution of the SPIRE-FTS bolometers at 205\,$\mu$m.  Line fluxes and the associated errors are listed in Table~\ref{tab:data}.  

\begin{figure}[t]
\centering
\epsscale{1.15}
\plotone{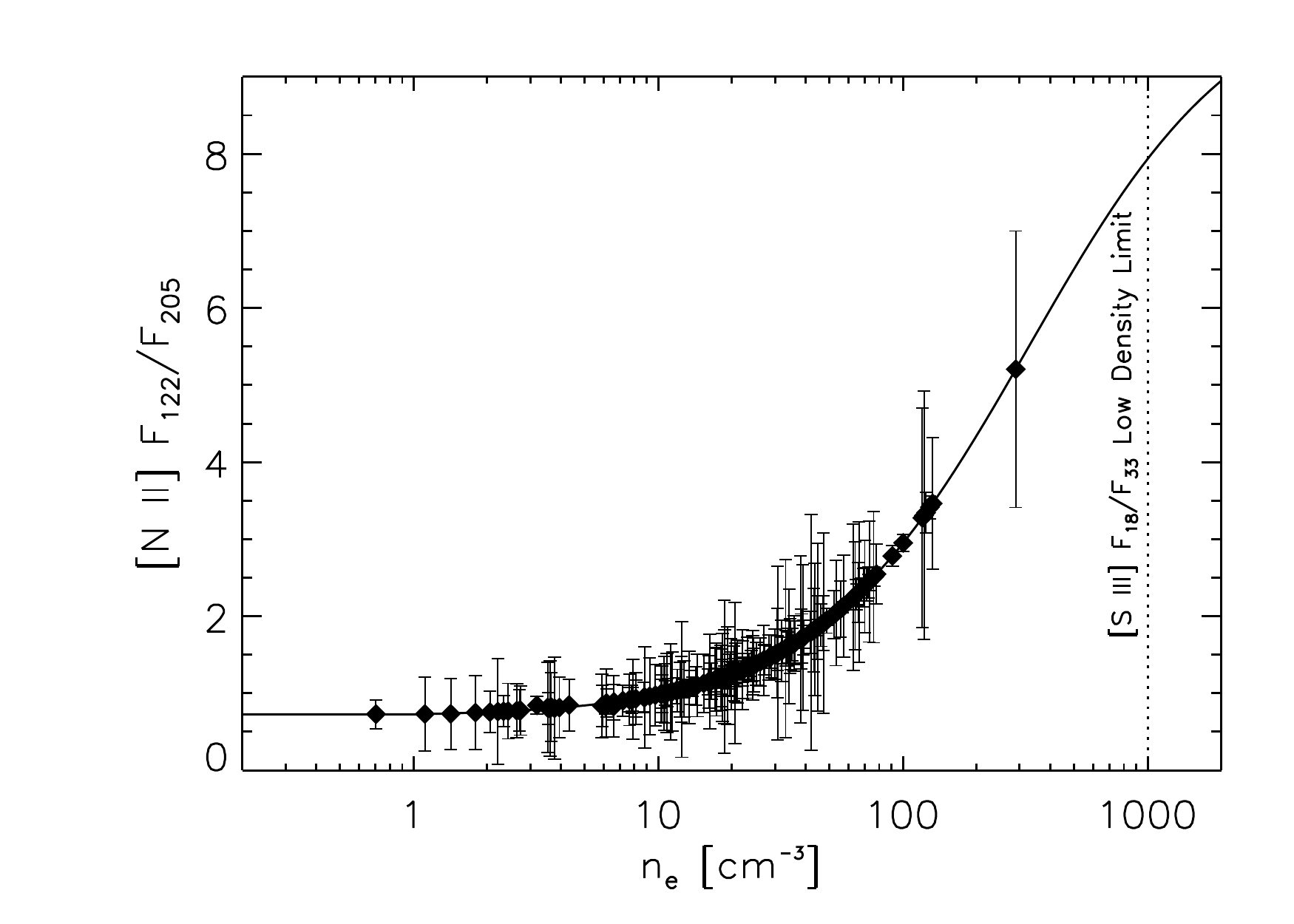}
\caption{The observed ratio of the two far-IR \nii\ lines plotted as a function of density, derived from the theoretical line ratio, shown as the solid line.  While the {\it Herschel} data are consistent with \nii\ originating in the low-density regime, as defined by [S\,\iii] emission (n$\leq$1000), N$^+$ ions clearly exist in a range of low-density environments, affecting the predicted ratio of \cii/\nii\,122\,$\mu$m.  This means that \cii/\nii\,122\,$\mu$m is an unreliable tracer of ionized \cii\, and observations of the \nii\,205\,$\mu$m is required to trace the contributions to \cii\ from ionized gas. }  
\label{fig:density}
\end{figure}

\subsection{Density and \cii\ Emission}
As previously noted, to use the \cii\,158\,$\mu$m line as a diagnostic for e.g., heating efficiency or star formation rate, it is necessary to distinguish the phases from which it arises.  Given the sensitivity of {\it Herschel}, one possible method to isolate the contribution of \cii\ emission from ionized gas is to use the \nii\,122\,$\mu$m line which is emitted purely from the ionized phase \citep{croxall2012,beirao2012}.  The ratio of \cii\,158\,$\mu$m to \nii\,122\,$\mu$m is, however, density dependent.  While the mid-IR [S\,\iii] lines are often used to determine the density of interstellar gas, the critical density of these lines limits their diagnostic use in very low density gas.  As shown in Figure \ref{fig:density} ionized gas clearly occupies a range of densities below the sensitivity of the \siii\ ratio  (i.e., $\leq$ 1000\,cm$^{-3}$); see also \citet{HerreraCamus2016} for additional discussion.  On the other hand, the critical densities of the \cii\,158\,$\mu$m and \nii\,205\,$\mu$m lines are very well matched, making their ratio relatively insensitive to density.  Thus, detections of the \nii\,205\,$\mu$m line from SPIRE-FTS enable us to determine a robust ionized fraction of the \cii\ emission independent of density.

\section{Results}
\begin{figure}[t]
\centering
\epsscale{1.15}
\plotone{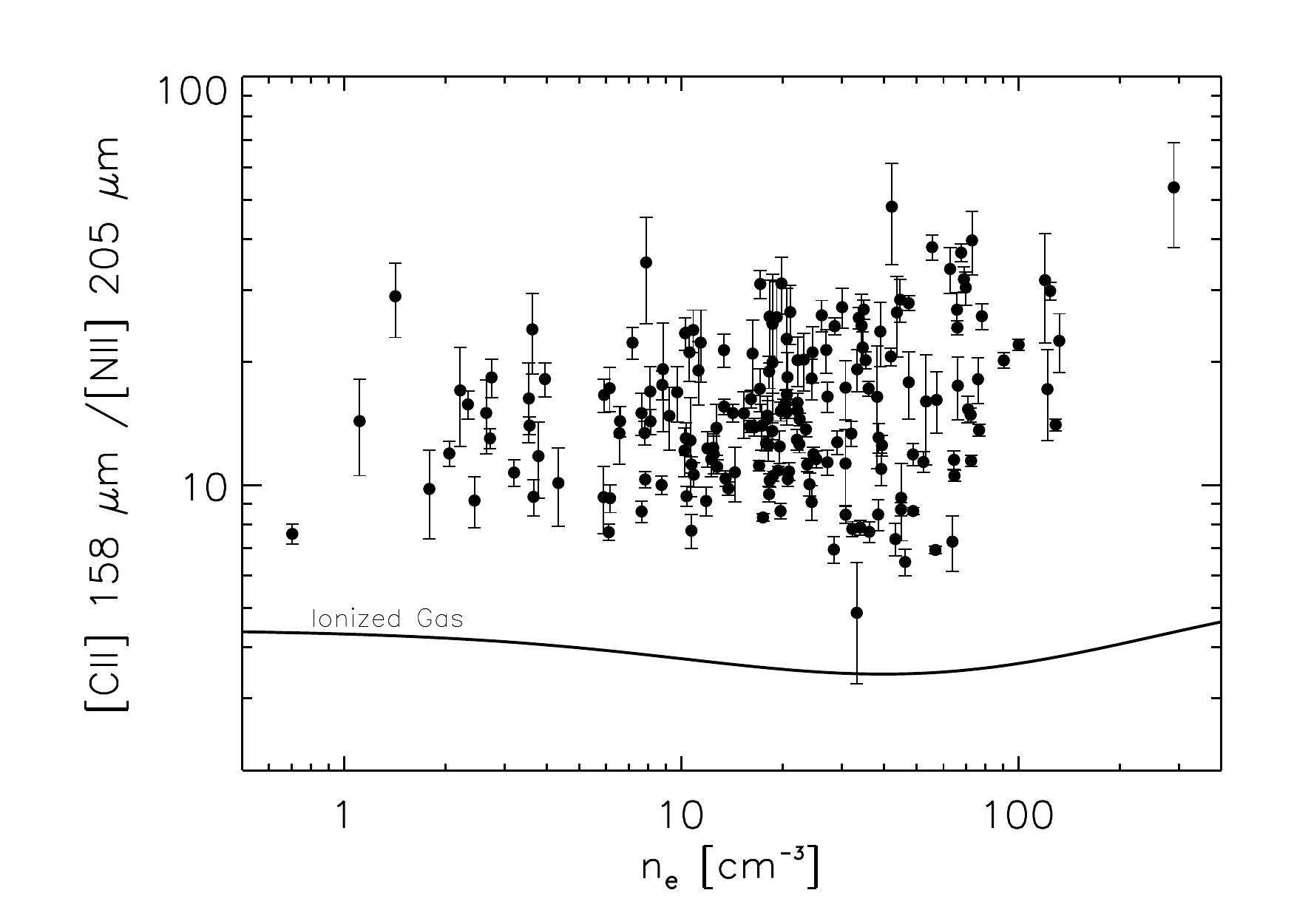}
\caption{The ratio of total [C\,\ii]/[N\,\ii]  205\,$\mu$m as a function of density.  Data show the line ratio observed in the bulk ISM of nearby galaxies, whereas the line denotes the theoretical predictions for ionized gas.  The fact that all ratios are elevated above this prediction indicates that non-ionized gas makes a significant contribution to the \cii\ emission.  The flatness of the predicted contribution from ionized gas demonstrates how can avoid the dependence  of \cii/\nii\ on density.}
\label{fig:model}
\end{figure}

In Figure \ref{fig:model} we plot the observed \cii/\nii\ ratios for regions with measurements in all three far-IR lines of interest (\cii\,158\,$\mu$m, \nii\,122\,$\mu$m, and  \nii\,205\,$\mu$m) as a function of the electron density derived from the \nii\,122/205 line ratio in Figure \ref{fig:density}.  The solid line indicates the predicted ratio of \cii/\nii\,205\,$\mu$m, if all of the \cii\ originated in the ionized gas phase.  Since all the observations are above these lines we clearly observe a substantial amount of \cii\ emission from the neutral gas.  We calculate the fraction of the \cii\ originating in the neutral ISM, f$_{\rm{[C\,II],Neutral}}$, by subtracting the contribution of \cii\ due to ionized gas,
\begin{equation}
	f_{\rm{[C\,II],Neutral}} = \frac{\rm{[C\,II]} - R_{ionized}\times \rm{[N\,II]} \,205\,\mu m} {\rm{[C\,II]}},
\end{equation}
where R$_{\rm{ionized}}$ is the \cii/\nii\ ratio in ionized gas derived using the collision rates of \citet{tayal2008}\footnote{We note that the adopted collision strengths do not differ significantly from the widely used values reported in \citet{blum1992}.} for \cii\ and \citet{tayal2011} for \nii\ and assuming Galactic gas phase abundances of carbon \citep[1.6$\times$10$^{-4}$ per hydrogen nucleus,][]{sofia2004} and nitrogen \citep[7.5$\times$10$^{-5}$ per hydrogen nucleus,][]{meyer1997}.  We remind the reader that $\rm{R_{ionized}}$ is the [CII]158um/[NII]205um ratio, $\approx4.0$, and is nearly independent of $n_e$.

\begin{figure*}[t]
\centering
\epsscale{0.57}
\plotone{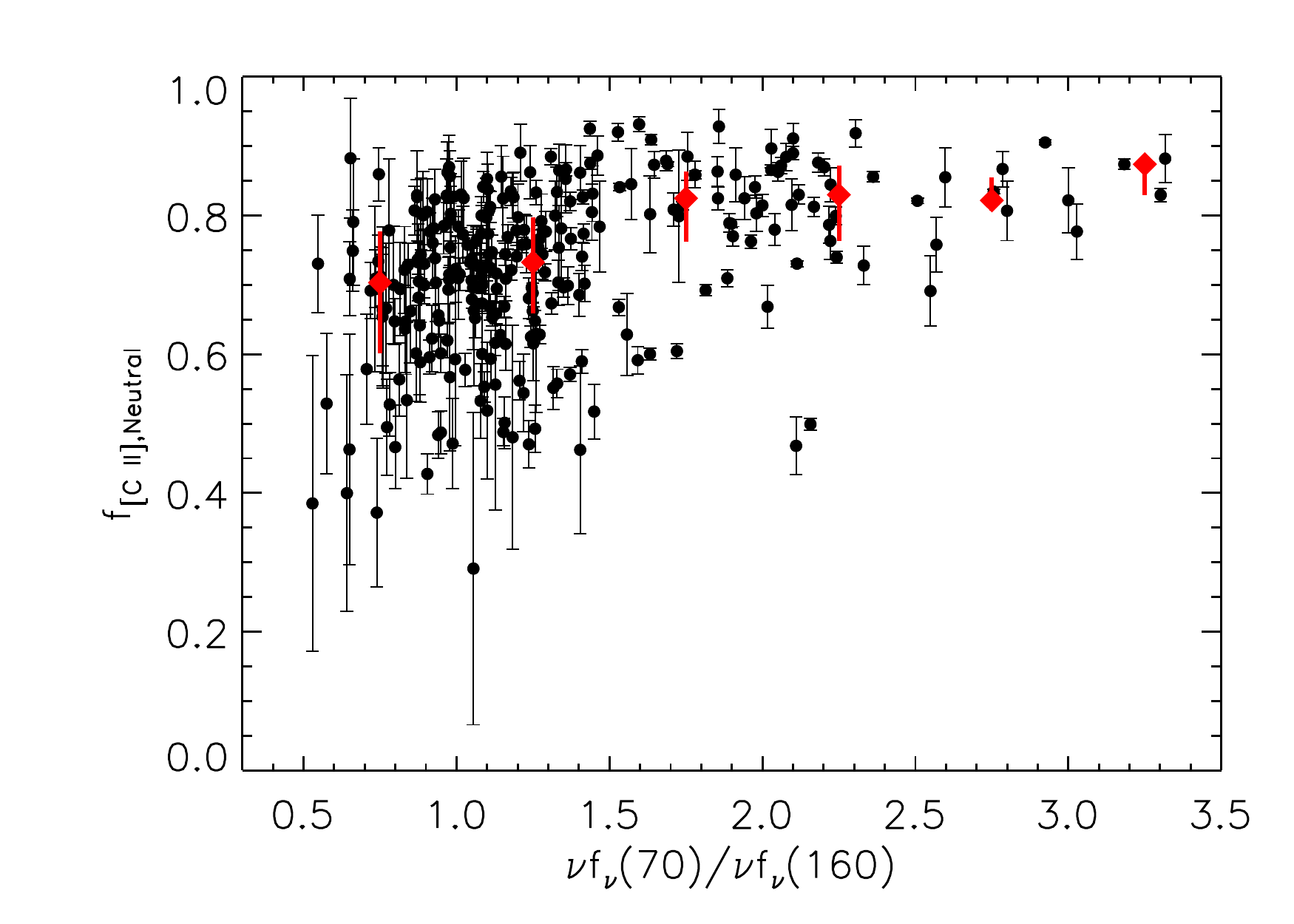}
\plotone{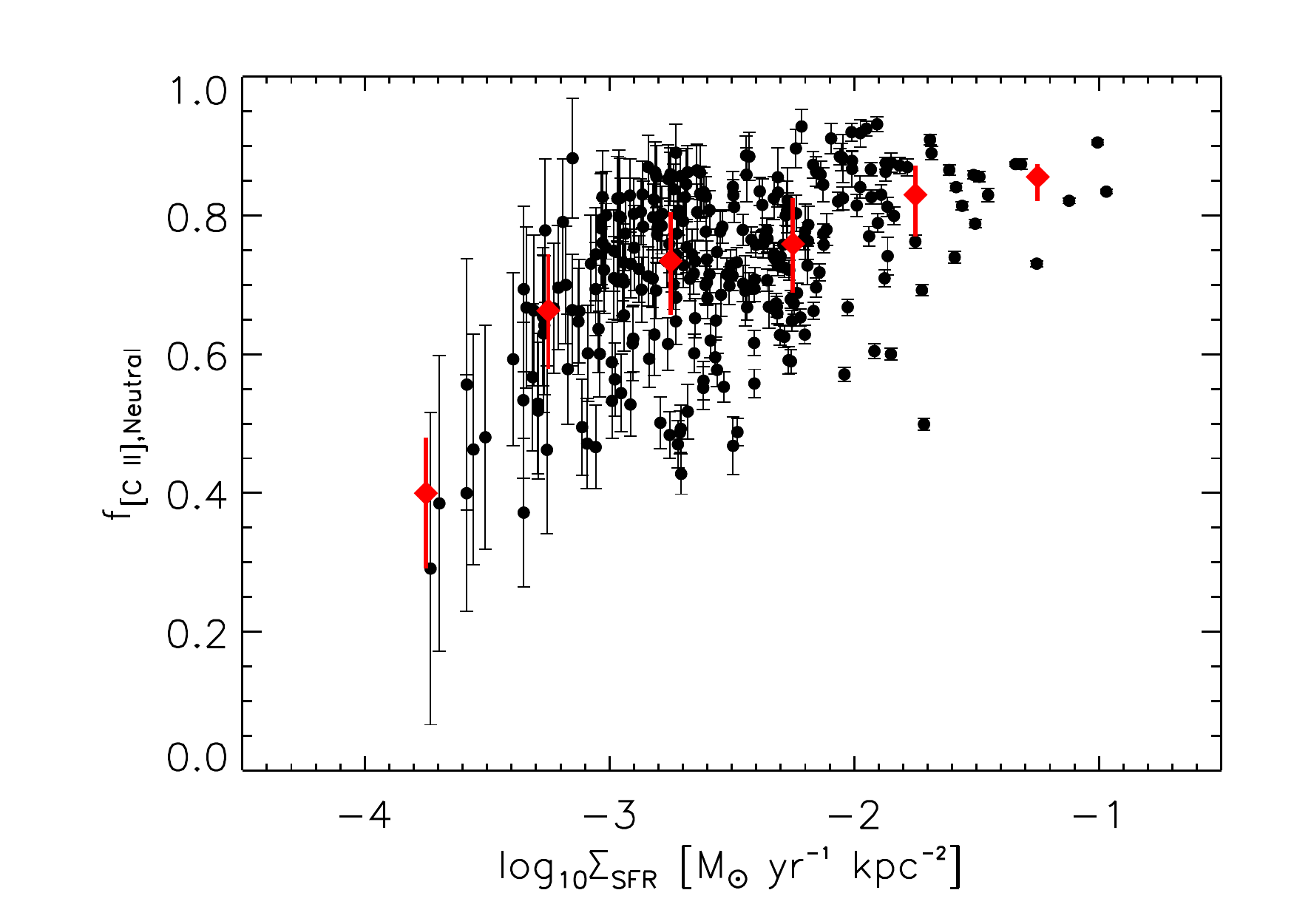}
\caption{The fraction of \cii\ emission that originates in the neutral phase, assuming a solar C/N ratio, plotted as a function of the far-IR color (left) and the star formation rate surface density (right).  Data binned to equal intervals are plotted as red diamonds, with error bars denoting the 25-75\% percentile ranges in each bin.  While there are weak correlations there is also significant scatter, particularly in gas described by a cool dust temperature.}  
\label{fig:fpdr}
\end{figure*}

We measure an average \cii/\nii\,205\,$\mu$m ratio of 17.5, with half of all regions lying between 11--22.  This indicates that (74$\pm$8)\% of \cii\ emission originates in the neutral phase of the ISM, with half of all regions characterized by neutral fractions that lie between 66--82\%; in very good agreement with \citet{rig13}.  

Possible changes in the neutral fraction have been a source of concern in understanding the deficit of \cii\ relative to continuum emission at warm IR colors \citep[e.g.][]{malhotra2001}.   
In Figure~\ref{fig:fpdr}, we show $f_{\rm{[C\,II],Neutral}}$ as a function of the far-infrared color, $\nu f_\nu$(70 $\mu$m)/$\nu f_\nu$(160 $\mu$m) (left), and the star formation rate surface density (right), calculated following the procedure of \citet{HerreraCamus2015}.  This reveals a rough trend such that regions exhibiting a warmer far-infrared color also tend to have a larger fraction of their \cii\ emission originating in the neutral phase of the ISM.  
The trend seen in Figure 3 might reflect increasing pressures in HII regions leading to suppression of both \nii\ and \cii.  If the \cii\ emission from neutral regions is suppressed less by pressure (\cii\ remains the dominant coolant in the neutral gas, so the gas temperature will rise until balanced by \cii\ emission) this could account for the observed tendency for $f_{[CII],neutral}$ to rise with increasing $\Sigma_{SFR}$ and 70/160 flux ratio.
        
{\it Herschel} observations of \cii\ emission in dwarf galaxies \citep{cormier2015} reveal elevated ratios of  \cii/\nii\,122\,$\mu$m, suggesting a small fraction of the \cii\ emission originates in ionized regions in low metallicity galaxies. And a velocity-resolved SOFIA/GREAT survey of low metallicity dwarf galaxy NGC\,4124 could associated only 9\% of \cii\ with cold neutral gas.

\begin{figure}[t] 
   \epsscale{1.15}
   \centering
   \plotone{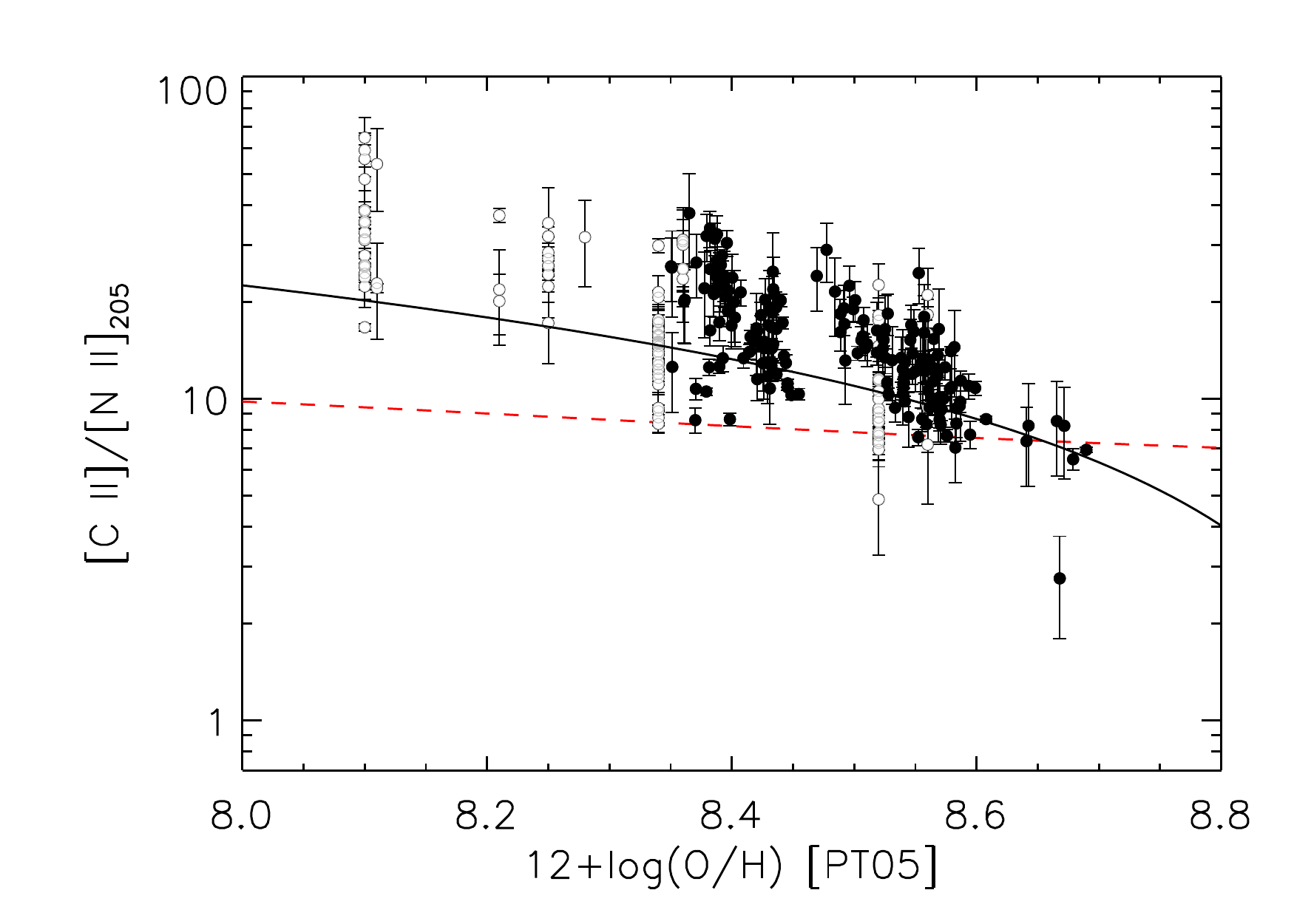}
   \plotone{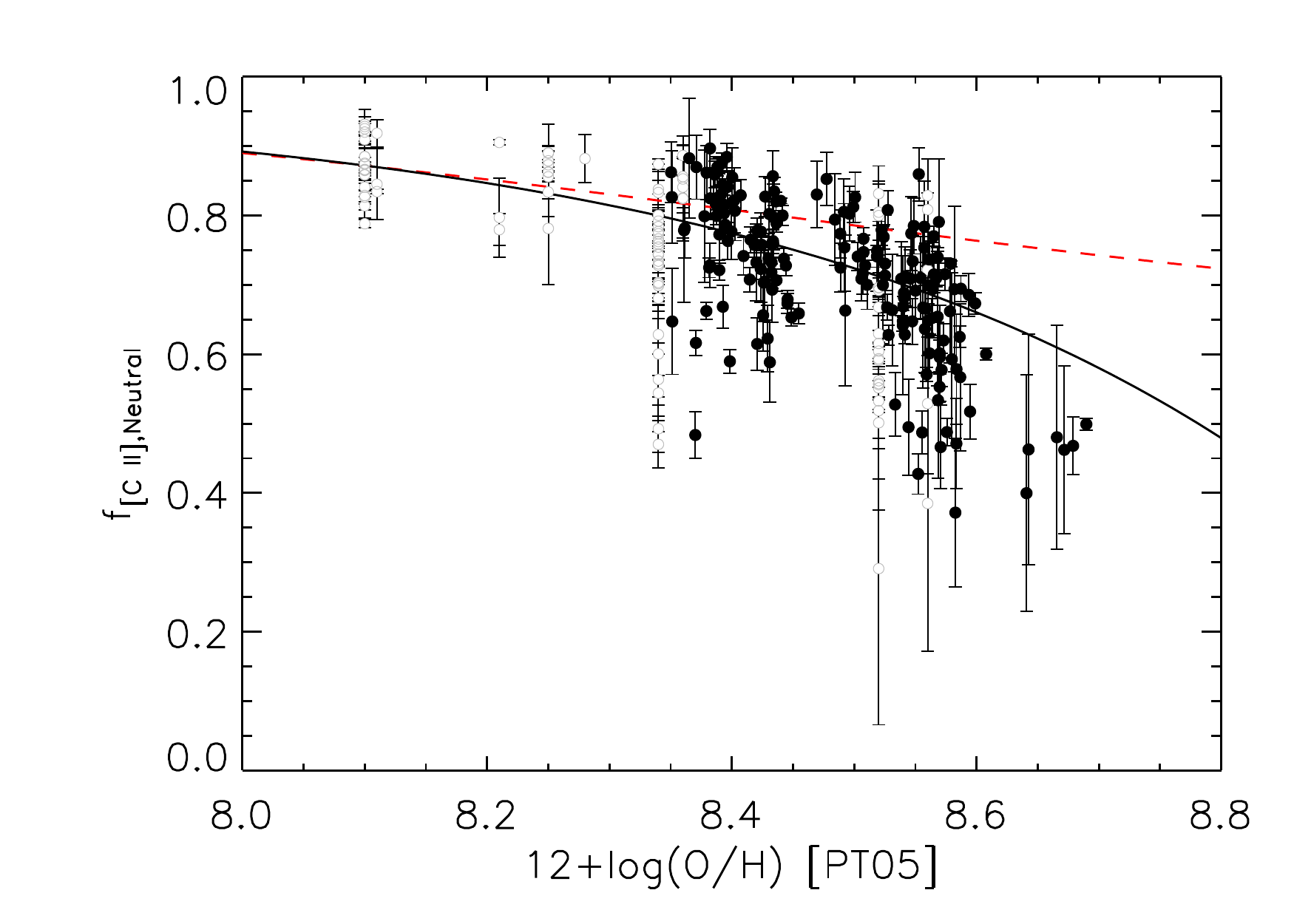}
   \caption{Top: The ratio of [C\,\ii]/[N\,\ii]\,205\,$\mu$m as a function of gas-phase oxygen abundance.  Bottom: The fraction of \cii\ emission originating in the neutral phase plotted as a function of the metallicity, assuming C/N$_\odot$.  
Regions in galaxies for which \citet{moustakas2010} determined a gradient are plotted as filled points while open points indicate global O/H measurements.  
The black lines represent a fit to $f_{\rm{[C\,II],Neutral}}$ and the linear O/H ratio while the dashed red lines show the arbitrarily normalized, modest trends expected if the \cii\ neutral fraction were in fact fixed, and changes in  [C\,\ii]/[N\,\ii]\,205\,$\mu$m and $f_{\rm{[C\,II],Neutral}}$ were driven solely by underlying C/N variation with 12+log(O/H) as expressed in Eq. 2.  The O/H abundance is on the scale of \citet{PT05}.}  
   \label{fig:fpdr_oh}
\end{figure}

Given that our data extend to higher metallicity, complementing the investigation of the neutral fraction of the \cii\ emission toward more metal-rich environments, we can better investigate the effects of metallicity on the origins of \cii\ emission.  Furthermore, by using the \nii\,205\,$\mu$m line we are not sensitive to systematic changes in density.  Adopting the oxygen abundances from \citet{moustakas2010} we plot the observed \cii/\nii\,205\,$\mu$m line ratio as a function of metallicity in the top panel of Figure \ref{fig:fpdr_oh}.  In galaxies for which metallicity gradients are available from \citeauthor{moustakas2010}, we determine the appropriate O/H for the location of the region; these are plotted as solid points.  For the rest of the galaxies where no abundance gradients are reported in the literature, we adopt a global O/H and plot them as open points.  There is a clear trend showing decreasing \cii/\nii\ ratio with increasing metallicity.   

Part of this trend may be the result of chemical evolution due to the star formation history of a galaxy.  As the metallicity of a system increases, the ratios of N/O and N/C also increase \citep[e.g.][]{Nieva2012}.   In particular, the N/O ratio exhibits a substantial increase once secondary production of nitrogen becomes important \citep{vilacostas}. 

To evaluate the impact caused by the variation of the C/N ratio as a function of the metallicity we calculate this ratio as a function of the O/H ratio using: 
\begin{equation}
  C/N = \frac{1.546 + 2.76\,(Z/Z_\sun)}{0.181 + 0.819\,(Z/Z_\sun)},
\end{equation}  
a functional fit based on stellar determinations for the variations with metallicity from \citet{Nieva2012} as adopted in the most recent version of the MAPPINGS code \citep{dopita2013}.  Although these trends track the general evolution of the C/N ratio, they do not take into account the significant scatter that is seen in \citet{Nieva2012}.  Nevertheless, the observed scatter is insufficient to explain the significant observed trend of \cii/\nii\ with metallicity.  We combine the C/N ratio as a function of O/H with the collision strengths of \citet{tayal2011} for \nii\ and \citet{tayal2008} for \cii\ to plot the expected evolution of the \cii/\nii\,205\,$\mu$m ratio as the red dashed lines in  Figure \ref{fig:fpdr_oh}.  The expected evolution of the C/N ratio with changing oxygen abundance produces a much flatter correlation than is observed.

Figure \ref{fig:fpdr_oh} shows that observations of the \cii/\nii\,205\,$\mu$m ratio provide some indication of gas-phase metallicity.  This was suggested as a possibility by the photoionization models employed by \citet{nagao2012}, who fit ALMA observations of a sub-millimeter galaxy, though this is the first time this trend has been seen in observations of local systems where the metallicity can be measured using standard methods \citep{agn3}.  We plot the linear fit to this empirical relation, 
\begin{equation}\rm[O/H]_{PT05} = 8.97 - 0.043 \times \rm{[C\,II]}\,158\,\mu \rm{m}/\rm{[N\,II]}\,205\,\mu \rm{m},\end{equation}
as a solid black line in the top panel of Figure \ref{fig:fpdr_oh}\footnote{The curvature in this line is due to the logarithmic scaling of the plot}.  We stress that this fit is (1) only an empirical fit and (2) only for the \citet{PT05} calibration of the metallicity scale.  Furthermore, the large scatter in \cii/\nii\ at any given O/H will lead to a large uncertainty in the determination of O/H if no other constraints are employed.

In the bottom panel of Figure \ref{fig:fpdr_oh} we show the derived quantity $f_{\rm{[C\,II],Neutral}}$ as a function of the gas-phase oxygen abundance.  We again include, as a red dashed line, the expected modest trend in \cii\,158\,$\mu$m/\nii\,205\,$\mu$m due to the changing C/N ratio.
\begin{equation}f_{\rm{[C\,II],Neutral}}  = 0.97 - 778\times\rm[O/H]_{PT05}.\end{equation}
While we here use the \citet{PT05} calibration of abundances, these results are independent of the adopted calibration; for instance, using the calibration of \citet{KK04} would shift the points but not affect the overall relationship.


\section{Summary \& Discussion}

Using spatially resolved data obtained from the PACS and SPIRE instruments on board the {\it Herschel} Space Observatory, we have analyzed the fraction of \cii\ emission originating in the ionized and neutral ISM of 21 nearby galaxies from the Beyond the Peak and KINGFISH programs.  We measure the ionized fraction of \cii\ by using the ratio of \cii\,158\,$\mu$m to \nii\,205\,$\mu$m, a method that has the unique advantage of being insensitive to density.  Our result that elevated levels of \cii/\nii\ are observed at lower metallicity is consistent with observations of dwarf galaxies \citep{cormier2015}. Similarly, these results are also consistent with recent investigations into the origin of Galactic plane emission carried out in \cii\ \citep{pineda2013,Velusamy2014} and \nii\ \citep{goldsmith2015} that find 1/3 -- 1/2 of Galactic \cii\ emission is associated with ionized gas.

In our sample of 21 resolved, nearby galaxies we find: 

\begin{itemize}

\item By directly accounting for the \cii\ associated with ionized gas, we find the typical fraction of \cii\ originating from neutral gas in our galaxy sample is 74\%, and half of all regions have \cii\ neutral fractions between 66\% and 82\%.

\item Weak correlations exist between the fraction of \cii\ originating in the neutral medium and the far-IR color  and $\Sigma_{\rm SFR}$, such that gas characterized by warm dust temperatures and/or high $\Sigma_{\rm SFR}$ has a smaller ionized fraction.  However, there is significant scatter, particularly in regions described by cool dust color-temperatures. 

\item A correlation is found between the measured \cii/\nii\,205\,$\mu$m and the gas phase oxygen abundance of gas,  such that metal poor galaxies have a significantly \emph{lower} fraction of their \cii\ emission arising from ionized regions, whereas very metal rich regions may have up to half of their \cii\ emission originating from the ionized phase.

\item The changes in ionized fraction we infer from trends in the observed \cii/\nii\,205\,$\mu$m ratio as a function of gas phase oxygen abundance are substantially larger than could be explained by the very mild modeled dependence of the underlying C/N abundance ratio on oxygen abundance itself.

\item The decrease of $f_{\rm{[C\,II],Neutral}}$ is better correlated with metallicity than other likely parameters such as FIR color, density, \cii\ luminosity, or star-formation surface density.  This observed decrease could be due to the harder radiation fields produced in a low-metallicity environments that may increase the relative abundance of C$^{++}$ in ionized gas at the expense of C$^+$, leading to an increased $f_{\rm{[C\,II],Neutral}}$. 

\end{itemize}

With the caveat that we cover only a modest range in gas phase metal content  (from 8.1--8.7; approximately 1/4--1 $Z_\sun$) and star formation rate density (well below the observed and theoretical starburst maximum near $10^3$\,M$_\sun $\,yr$^{-1} $\,kpc$^{-2}$), we can highlight several implications of these results:

\begin{itemize}
\item While \cii\ does inhabit both neutral and ionized regions within galaxies, the bulk of \cii\ cooling arises from dense or diffuse neutral gas.
  
\item 
  The low overall ionized \cii\ fraction in galaxies implies that PDR models which assess density and radiative heating intensity using this line can be employed with only modest concern of the effects of ionized contributions, although ignoring ionized \cii\ emission altogether could introduce biases at lower star formation rate densities and higher metallicity (where the ionized fraction peaks).

\item Given the tight scaling between molecular gas surface density and star formation rate density \citep[e.g.][]{kennicutt2007}, and the dependence of the \cii\ deficit and total surface brightness on the latter \citep{HerreraCamus2015,smith2016}, a correlation between \cii\ intensity and molecular content could be anticipated.  Indeed, \cii\ is sometimes considered as a potential direct tracer of molecular gas in galaxies. Although we cannot here discriminate between molecular and atomic phases (with CO-dark gas contributing substantially to \cii\ in some environments), the uncertainty introduced by varying ionized gas contributions inherent in any mapping between \cii\ emission and molecular gas surface density must be substantially less than a factor of 2, becoming negligible at high star formation rate surface densities.  

\item 
  The fact that the fraction of \cii\ arising from neutral gas rises to near unity as metallicities drop below 1/4 solar means that the high fractional \cii/TIR luminosities found in many star-forming low metallicity galaxies (in contrast to the deep \cii/TIR deficits seen in ULIRGs and other compact star-forming systems) cannot be attributable to additional contributions to the line from energetic ionized regions.
  
\end{itemize}

\acknowledgments
KVC would like to thank T. Diaz-Santos, G. Accurso, R. Pogge, \& P. Goldsmith for useful discussions regarding this work.  
KVC acknowledges support from the Deutsche Forschungsgemeinschaft Priority Program ISM-SPP 1573 and the MPIA and thanks the KINGFISH \& BTP teams for their support.  JDS acknowledges visiting support from the Alexander von Humboldt Foundation and the MPIA.  BG  acknowledges the support of the Australian Research Council via a Future Fellowship (FT140101202).  ADB acknowledges partial support from NSF-AST0955836 and 1412419.  BTD aknowledges partial support from NSF-AST1408723.
Beyond the Peak research has been supported by a NASA/JPL grant (RSA 1427378). 
HIPE is a joint development by the Herschel Science Ground Segment Consortium, consisting of ESA, the NASA Herschel Science Center, and the HIFI, PACS and SPIRE consortia.  
This work is based on observations made with Herschel, a ESA Cornerstone Mission with significant participation by NASA. Support for this work was provided by NASA through an award issued by JPL/Caltech.  
This research has made use of the NASA/IPAC Extragalactic Database which is operated by JPL/Caltech, under contract with NASA.

\end{document}